\documentclass[reprint,pra]{revtex4-1}
\usepackage{amsfonts}
\usepackage{amssymb}
\usepackage{amsmath}
\usepackage{hyperref}
\usepackage{graphicx}
\usepackage{epstopdf}
\usepackage{subfig}
\usepackage{makeidx}
\usepackage[page,toc,title,titletoc]{appendix}

\newcommand{\vk}{\ensuremath{\mathbf{k}}}

\newcommand{\vp}{\ensuremath{\mathbf{p}}}
\providecommand{\sch}{{Schr\"{o}dinger }}
\newcommand{\nth}[1]{\ensuremath{\frac{1}{#1}}}

\newcommand{\br}[1]{\ensuremath{\left(#1\right)}}
\newcommand{\mbr}[1]{\ensuremath{\left[#1\right]}}

\providecommand{\abs}[1]{\ensuremath{\left\lvert{#1}\right\rvert}}

\newcommand{\mtrx}[1]{\ensuremath{\begin{pmatrix}#1\end{pmatrix}}}

\newcommand{\dg}{\ensuremath{\dagger}}
\newcommand{\tr}{\ensuremath{\text{tr}}}
\newcommand{\nG}{\ensuremath{\hat{\mathcal{G}}^{-1}}}

\newcommand{\av}[1]{\ensuremath{\bigl<{#1}\bigr>}}

\newcommand{\ket}[1]{\ensuremath{\left|#1\right>}}

\begin{document}

\title{BEC-BCS Crossover with Feshbach Resonance for a Three-Hyperfine-Species Model}
\author{Guojun Zhu}
\email{zhu.guojun2010@gmail.com}
\affiliation{Department of Physics, University of Illinois at Urbana-Champaign}
\author{Anthony J. Leggett}
\affiliation{Department of Physics, University of Illinois at Urbana-Champaign}
\email{alleggett@illinois.edu}

\begin{abstract}
We consider the behavior of an ultracold Fermi gas across a narrow
Feshbach resonance, where the occupation of the closed channel may not be
negligible. While the corrections to the single-channel formulae associated
with the nonzero chemical potential and with particle conservation have
been considered in the existing literature, there is a further effect, namely the
``inter-channel Pauli exclusion principle" associated with the fact that
a single hyperfine species may be common to the two channels. We focus on
this effect and show that, as intuitively expected, the resulting
corrections are of order $E_F/\eta$, where $E_F$ is the Fermi energy of the gas
in the absence of interactions and $\eta$ is the Zeeman energy difference
between the two channels. We also consider the related corrections
to the fermionic excitation spectrum, and briefly discuss the collective
modes of the system.
\end{abstract}
\maketitle

\section{Introduction}

In a low-temperature dilute system, a short-range interaction can be characterized with a single parameter, $a_s$, a.k.a. the s-wave scattering length.    A very desirable property of the Feshbach resonance in such a system is that the effective interaction is tunable experimentally through the Zeeman energy difference between channels which is in turn  tunable through  instruments such as a magnetic field  \cite{Fano,nuclear,ChinRMP,Pethick}. 
 \begin{equation}
a_{s}(B)=a_{bg}\br{1+\frac{\Delta{B}}{B-B_{0}}}
\end{equation}
where $B_{0}$ is the magnetic field at which the $a_{s}$ diverges, i.e., the  resonant point. 
This unique ability gives physicists a rare opportunity to study  a many-body system under various interaction strengths,  and thus connect different physics originally developed separately.  Particularly for the fermionic gas, there are a series of  theoretical works about uniform treatment of  BEC and BCS since the 1960s \cite{Eagle,LeggettCrossover,Nozieres,RanderiaBEC}, for which the dilute ultracold fermionic alkali gas with  Feshbach resonances provides a perfect testing ground.  Indeed,  these theories work quite well  qualitatively.  


\begin{figure}[htbp]
\begin{center}
\includegraphics[width=0.8\columnwidth]{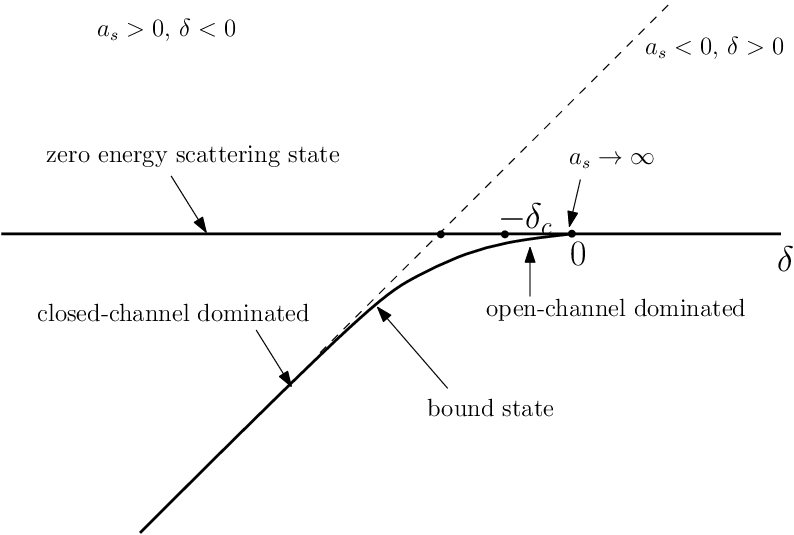}
\caption{Energy levels in a Feshbach resonance\label{fig:intro:levels}} 
\parbox{0.9\columnwidth}{\raggedright \small $\delta$ is the energy detuning from the resonance point.  The horizontal line stands for the zero energy s-wave scattering state, $\psi\sim\nth{r}-\nth{a_s}$, which exists for any detuning.  The lower solid line stands for the real bound state, which only exists for negative detuning ($\delta<0$, $a_s>0$). The dashed line stands for the (uncoupled) closed-channel bound state.  An interesting point to notice is that the real bound state appears earlier than the cross point of the (uncoupled) closed-channel bound-state level and zero energy. Another important point to notice is the negative detuning $-\delta_c$.  When the negative detuning is smaller than $\delta_c$, this real bound state is composed mostly with atoms in the open channel and vice versa.  
}

\end{center}
\end{figure}

  The two-body theory of the Feshbach resonance has a characteristic  parameter, $\delta_c\sim{}a_{bg}^2(\Delta{B})^2$, defined as  the detuning energy at which the weight of the bound state shifts from predominantly in the open channel to predominantly in the closed channel (see Fig. \ref{fig:intro:levels}) \cite{Leggett}.  Na\"{i}vely speaking, on the negative detuning side of any resonance (i.e. $\delta<0$), the two particles should mostly stay  in a (virtual) bound state of the closed channel (or ``virtual state'' in some other-type resonances).  However, at the resonance point  of a Feshbach resonance ($a_s\to\pm\infty$), the atoms are mostly still in the open channel, and they do so down to a negative detuning $\delta\sim-\delta_c$. Only when the negative detuning from resonance is much more  than $\delta_c$, do atoms have the majority weight in the closed channel.    
  
  When considering a many-body system with a Feshbach resonance, an important question is how this energy scale, $\delta_c$, compares to the  many-body energy scale, namely, the Fermi energy of the free fermionic atoms, $E_F$. In the region not too far away from the resonance ($\abs{\delta}\ll\delta_{c}$), the closed-channel weight is negligible if the Fermi energy is much smaller than $\delta_c$, (i.e., \emph{broad resonance}).  Crossover experiments are usually performed at detuning not too far from the resonance, and hence the closed channel can be safely ignored at the many-body level. Eventually, when the detuning is too far away, $\abs{\delta}\gg\delta_{c}$, the bound state is almost like the    uncoupled closed-channel bound state with a little dressing from the open channel.  
  Nevertheless, such a large detuning in the broad resonance is not very interesting because the  resonance effect is very small and the s-wave scattering length, $a_s$, is close to its background  value then.  
 Given the above consideration, for most purposes, we can almost neglect the closed channel  and the problem can be well-described as a two-species fermion system with a tunable interaction when it is not too far away from the resonance.  The Feshbach resonance indeed serves as a simple ``magic'' knob to change the interaction strength.  The original  theories developed on  single-channel models  apply to this case directly.  This is also the situation for two  popular experimental systems (${}^{6}\text{Li}$ atoms at 834G, $^{40}\text{K}$ atoms at 224G).   Many theoretical works have been developed using either the single-channel model or  the two-channel model with the broad resonance assumption (e.g. \cite{Holland01,HoUniversal,Fuchs04}). On the contrary, when the Fermi energy of the free fermionic atoms is  comparable to or even larger than $\delta_c$, the closed channel has to be included at the many-body level even for small detuning. Such a situation, previously considered in some works \cite{JacksonNarrow,GurarieNarrow}, is the focus of the current work.
  

To complicate the problem  further,   configurations of Feshbach resonances often have one common hyperfine species between the two channels. There are three hyperfine species in the  two channels instead of four species (two for each channel).  Two most common systems (${}^{6}\text{Li}$ at 834G, $^{40}\text{K}$ at 224G) both contain three species of fermions although they are broad resonances.  The Pauli exclusion principle prevents  atoms of both channels from occupying the same momentum level simultaneously because of this common species.  This ``inter-channel Pauli exclusion'' has no counterpart in the two-body physics. This peculiar effect  in many-body crossover problems has  received little theoretical attention up to now.    Nevertheless,   narrow resonances do exist \cite{ChinRMP} and it is not  inconceivable to perform many-body experiments using such resonances.  The central concern of this paper is about these situations. 

Roughly speaking, turning from two-body systems with Feshbach resonances to many-body systems brings three new effects into the original two-body problem.  The first effect is closely associated with the Fermi energy:  For a many-body fermionic system at low temperature, most fermions are inactive; only the fermions close to the Fermi surface participate in the interaction processes. Therefore, the energy often needs to be measured from the Fermi surface instead of from zero as in a two-body situation. 
The second effect relates to  particle conservation. Unlike in the single-channel problem, there are two relevant densities in the two-channel problem: the density of atoms in the open channel, $n_{o}$, and the density of atoms in the closed channel, $n_{c}$. When the closed-channel weight is small (broad resonance), it is legitimate to treat the total density as the same as the open-channel density.  However, in the narrow resonance, where the closed-channel weight is not negligible, counting becomes complicated.  Extra care is required to specify which channel quantities such as ``density'' refer to.  These two aspects have been   extensively studied previously \cite{JacksonNarrow,GurarieNarrow}.

The last effect is unique to the three-species problem, where one common species is shared by both channels.  The phase spaces of two channels overlap because of  the common species, which prevents both channels have the occupation in the same  level simultaneously. This effect is controlled by the wave-function overlap of the states in the two channels. A rough estimate of this overlap can be made: The uncoupled closed-channel bound state which is in resonance with the open-channel zero energy threshold has  relatively small  spatial extension, $a_c$.  Its binding energy $E_b$ is close to the Zeeman energy difference between two channels, $\eta$, when considering only situations not far from resonances.  On the other hand, fermions in the open channel fill the lowest  momentum states up to typically the Fermi energy, $E_F$.  By a simple dimensional argument, the ratio $E_F/\eta$ must control the overlap effect.   How it modifies the many-body picture is the central topic of this paper. 


The present paper is divided as follows:
Section \ref{sec:model}  defines the many-body model and introduces the appropriate notation for the 3-species case. Section \ref{sec:mean} lists the mean-field result and its renormalization result. Section \ref{sec:fermionic} discusses the fermionic excitation within the mean-field level, and  section \ref{sec:bosonic} discusses the bosonic modes that are  beyond the mean-field level. We conclude and discuss our approach in Section \ref{sec:conclusion}.  A couple of detailed calculations are list in the Appendix and further details of the calculations may be found in ref. \cite{Zhuthesis}. 

\section{Model and notation\label{sec:model}}
We denote the three hyperfine species as $a$, $b$ and $c$, where the open channel contains the pair of species $(a,b)$ and the closed channel contains the pair of species $(a,c)$.  We denote the Grassmann variable for species $i$ by $\psi_i$ and write the  two channels in the notation:
\begin{equation}
(\bar\psi\bar\psi)=\mtrx{\bar\psi_{a}\bar\psi_{b}&\bar\psi_{a}\bar\psi_{c}}
\qquad(\psi\psi)=\mtrx{\psi_{b}\psi_{a}\\\psi_{c}\psi_{a}}
\end{equation}
The two-body interaction can then be written as a ($2\times2$)  hermitian  matrix  $\tilde{U}$ 
\begin{equation}
\tilde{U}\equiv{}\mtrx{U&Y\\Y^{*}&V}
\end{equation}
 Note that we have restricted the Hilbert space to only include pairs of $(a,b)$ and $(a,c)$.  When only close-to-resonance region is considered, this is a good approximation instead of  the most generic description, where all three species are treated equivalently.  Within this approximation, species $b$ and $c$ share the same chemical potential, but not with $a$; in other words, $n_a=n_b+n_c$.  
  We can now write the finite-temperature action as 
\begin{multline}\label{eq:pathInt2:actionFermi}
S(\bar\psi,\psi)=\int^{\beta}_{0}d\tau\int{d^{d}r}\\
\mbr{\sum_{j}\bar\psi_{j}(\partial_\tau-\nth{2m}\nabla^{2}-\mu+\eta_{j})\psi_{j}
-(\bar\psi\bar\psi)\tilde{U}(\psi\psi)}
\end{multline}
Here $\eta_{j}$ is the Zeeman energy for hyperfine species $j$. We choose the zero so that $\eta_{a}=\eta_{b}=0$, $\eta_{c}=\eta$.    

Let us illustrate above discussion in  one example.  In a common experimental setup for $^{6}$Li, atoms are usually  prepared in the two lowest hyperfine levels: described by the  direct product of two states, $\ket{F=\nth{2},F_{z}=-\nth{2}}\otimes\ket{F=\nth{2},F_{z}=+\nth{2}}$.  This is a good approximation until the two atoms are very close.  For the atom-atom interaction that  conserves the z-component of the total angular momentum, $F_{z}^{(1)}+F_{z}^{(2)}$, this channel mixes with four other possible channels of the same total z-direction angular momentum, i.e. $F_{z}^{(1)}+F_{z}^{(2)}=0$: $\ket{\nth{2},-\nth{2}}\otimes\ket{\frac{3}{2},+\nth{2}}$, $\ket{\frac{3}{2},-\nth{2}}\otimes\ket{\nth{2},+\nth{2}}$, $\ket{\frac{3}{2},+\frac{3}{2}}\otimes\ket{\frac{3}{2},-\frac{3}{2}}$, $\ket{\frac{3}{2},+\frac{1}{2}}\otimes\ket{\frac{3}{2},-\frac{1}{2}}$ (All states are labeled as $\ket{F,F_{z}}$).  Various resonances can take place. Nevertheless, it is usually sufficient to consider only the  one in resonance and neglect all others, when close to the resonance.  The closed channel in the most studied resonance with a magnetic filed close to 834G, is approximately $\ket{\frac{3}{2},-\nth{2}}\otimes\ket{\nth{2},+\nth{2}}$ and the resonance is a three-species resonance\cite{ZhangThesis,ChinRMP}. In this case, $a$ is  $\ket{\nth{2},+\nth{2}}$; $b$ is $\ket{-\nth{2},+\nth{2}}$; and $c$ is $\ket{\frac{3}{2},-\nth{2}}$.

We  perform the Hubbard-Stratonovich transformation on Eq.\ref{eq:pathInt2:actionFermi}.   Introduce 2-component  auxiliary fields (order parameters), $(\Delta_{1},\Delta_{2})$, coupled to the fermionic fields as 
\begin{equation}\label{eq:pathInt2:DeltaPhi}
\Delta\longrightarrow\Delta-\tilde{U}(\psi\psi)
\end{equation}
We can introduce  a spinor representation   
\begin{equation}
\bar\Psi=\mtrx{\bar\psi_{a}&\psi_{b}&\psi_{c}}\qquad\Psi=\mtrx{\psi_{a}\\\bar\psi_{b}\\\bar\psi_{c}}
\end{equation}
The action can then be rewritten in a more compact form with respect to $\Psi$ and $\bar\Psi$
\begin{equation}\label{eq:pathInt2:actionMixCompact}
S(\bar\Delta,\Delta,\bar\psi_{i},\psi_{i})=\int^{\beta}_{0}d\tau\int{d^{d}r}
	\mbr{\Delta^{\dg}\tilde{U}^{-1}\Delta-\bar\Psi\mathcal{G}^{-1}\Psi}
\end{equation}
where the fermionic correlation $\mathcal{G}^{-1}$ in the momentum-frequency representation is 
\begin{equation}\label{eq:nG}
\mathcal{G}^{-1}=
\begin{pmatrix}
i\omega_{n}-\xi_{\vk}&\Delta_{1}&\Delta_{2}\\
\bar\Delta_{1}&i\omega_{n}+\xi_{\vk}&0\\
\bar\Delta_{2}&0&i\omega_{n}+\xi_{\vk}+\eta
\end{pmatrix}
\end{equation}
where $\xi_{\vk}=\hbar^{2}k^{2}/2m-\mu$. 
The action in Eq. \ref{eq:pathInt2:actionMixCompact} is  bilinear in the quantities $\Psi$, $\bar\Psi$ and we can formally integrate them out, with the result 
\begin{equation}\label{eq:pathInt2:actionD}
S(\bar{\Delta},\Delta)=\int{dx}\br{\bar{\Delta}\tilde{U}^{-1}\Delta-\tr\ln\nG}
\end{equation}

\section{Mean-field result and renormalization\label{sec:mean}}
 Eq. (\ref{eq:nG}) can be inverted to get $G$.   The final mean-field equations are (for simplicity,  both $\Delta_{i}$'s are taken as real.\footnote{\label{foot:pathInt2:real}When $Y$ is not real, $\Delta_{1}$ and $\Delta_{2}$ cannot be both real even at the mean field level.  Nevertheless, we can require one  real, then the other will have a phase just to compensate the phase in $Y$.  The final conclusion can be verified to remain valid.   }) 
  \begin{equation}\label{eq:pathInt2:mf}
\mtrx{\Delta_1\\\Delta_2}=\mtrx{U&Y\\Y^{*}&V}\sum_{\vk}\mtrx{h_{1\vk}\\h_{2\vk}}
\end{equation}
  where $ h_{1\vk}$ and $ h_{2\vk}$ are expectations of the abnormal Green's function.
  \begin{gather}
  h_{1\vk}=\av{\psi_{a,-{\vk}}\psi_{b,+{\vk}}}
  =\Delta_{1}\frac{E_{1\,\vk}+\xi_{\vk}+\eta}{(E_{1\,\vk}+E_{2\,\vk})(E_{1\,\vk}+E_{3\,\vk})}\label{eq:pathInt2:h1}\\
  h_{2\vk}=\av{\psi_{a,-{\vk}}\psi_{c,+{\vk}}}
  =\Delta_{2}\frac{E_{1\,\vk}+\xi_{\vk}}{(E_{1\,\vk}+E_{2\,\vk})(E_{1\,\vk}+E_{3\,\vk})}\label{eq:pathInt2:h2}
  \end{gather}
   where  $E_{i\,\vk}$'s are the eigenvalues of the fermionic correlation Eq. \ref{eq:nG} (see details in Sec. \ref{sec:fermionic}). 
   
   There is one  number equation for each channel,  
\begin{gather*}
\sum_{\omega_{n}, \vk}G_{22}e^{(-i\omega_n\delta_-)}=N_{open}\\
\sum_{\omega_{n},\vk}G_{33}e^{(-i\omega_n\delta_-)}=N_{close}
\end{gather*}
 The Matsubara summation can be performed by the normal trick of multiplying the summand by a Fermi function and deforming the contour \footnote{see sec. 4.2.1 in \cite{Altland}, sec. 25 in \cite{Fetter}}.  For the summation at zero temperature, we just need to consider the positive roots, $E_{1\,\vk}$.  It is straightforward to find 
\begin{gather}
N_{\text{open}}=\sum_{\vk}\frac{(E_{1\,\vk}-\xi_{\vk})(E_{1\,\vk}+\xi_{\vk}+\eta)-\Delta_2^2}{(E_{1\,\vk}+E_{2\,\vk})(E_{1\,\vk}+E_{3\,\vk})}
\label{eq:pathInt2:numOpen}\\
N_{\text{closed}}=\sum_{\vk}\frac{(E_{1\,\vk}-\xi_{\vk})(E_{1\,\vk}+\xi_{\vk})-\Delta_1^2}{(E_{1\,\vk}+E_{2\,\vk})(E_{1\,\vk}+E_{3\,\vk})}
\label{eq:pathInt2:numClose}
\end{gather}
 \subsection{The Bogoliubov canonical transformation and the  fermionic excitation\label{sec:fermionic}}
 The fermionic correlation (Eq. \ref{eq:nG}) can be diagonalized with a (unitary) Bogoliubov canonical transformation.  We  break down the transformation into two steps $T_{\vk}$ and $L_{\vk}$ . 
\begin{equation}\label{eq:pathInt2:B}
B_{\omega_{n},\vk}=L_{\vk}^{\dg}T_{\vk}^{\dg}G_{\omega_{n},\vk}^{-1}T_{\vk}L_{\vk}
\end{equation} 
Here $B_{\vk}$ is the diagonal matrix; $T_{\vk}$ and $L_{\vk}$ are both unitary transformations.  We take $T_{\vk}$ as the canonical transformation at the broad resonance limit, i.e., when we can ignore the inter-channel Pauli exclusion. 
\begin{equation}\label{eq:pathInt2:T}
T_{\vk}=\mtrx{u_{\vk}&v_{\vk}&0\\-v_{\vk}&u_{\vk}&0\\0&0&1}
\end{equation}	
where $u_{k}$ and $v_{k}$ are defined in a similar fashion as in the single-channel BCS  problem
\begin{gather}
v_{\vk}^{2}\equiv1-u_{\vk}^{2}\equiv\nth{2}\br{1-\frac{\xi_{\vk}}{E_{\vk}}}\\
E_{\vk}\equiv(\xi_{\vk}^{2}+\Delta_{1}^{2})^{1/2}
\end{gather}
 Note that in the narrow resonance here,  $v_{\vk}^{2}$  does not carry the physical meaning of the occupation number of the (open-channel) atoms, and $E_{\vk}$   does not represent the fermionic excitation spectrum. $T_{\vk}$ can be taken as the direct sum of two parts, the first two rows/columns describe the open-channel excitations, while the third row/column describes the closed-channel ones.  It nevertheless is not sufficient to diagonalize the fermionic correlation in the narrow resonance.  

Introduce a dimensionless scale $\zeta$, (Please refer to Appendix  \ref{sec:diagonalize} and \ref{sec:pathApp:consistency} for details.)
\begin{equation}\label{eq:pathInt2:zetaDef}
\boxed{\zeta=\frac{\Delta_{2}^{2}}{\Delta_{1}\eta}\sim\br{\frac{E_F}{\eta}}^{\nth{2}}\ll1}
\end{equation}
Here both $\Delta_{1}$ and $\Delta_{2}$ are their mean-field (saddle point) values.
It is not hard to find  the additional unitary transformation $L_{\vk}$ to the   first order in $\zeta$ . (Please refer to Appendix \ref{sec:diagonalize} for details.)
\begin{gather}\label{eq:pathInt2:L1}
L_{\vk}\approx{}I+
\mtrx{0&-\frac{\Delta_{1}{}\Delta_{2}{}}{4E^{2}_{\vk}}&u_{\vk}\\
\frac{\Delta_{1}{}\Delta_{2}{}}{4E^{2}_{\vk}}&0&v_{\vk}\\
-u_{\vk}&-v_{\vk}&0
}\frac{\Delta_{2}{}}{\eta}
\equiv{}I+\delta_{k}\\
L^{\dg}_{\vk}=I-\delta_{\vk}\nonumber
\end{gather}
So we finally arrive at the diagonal matrix $B_{\omega_{n},\vk}$ to first order in $\zeta$
\begin{equation}\label{eq:pathInt2:Bapprox}
B_{\omega_{n},\vk}=i\omega_{n}I-
	\begin{pmatrix}E_{1}{}_{\vk}&0&0\\0&-E_{2}{}_{\vk}&0\\0&0&-E_{3}{}_{\vk}\end{pmatrix}
\end{equation}
The eigenvalues of $\nG$  in $B_{\omega_{n},\vk}$ describes the dispersion spectrum of the  fermionic excitations
\begin{align}\label{eq:pathInt2:xiExpand}
E_{1\vk}&\approx{}E_{\vk}+\frac{\Delta_{2}^{2}u_{\vk}^{2}}{\xi_{\vk}+\eta}
\approx{}E_{\vk}+u_{\vk}^{2}\Delta_{1}\zeta\\
E_{2\vk}&\approx{}E_{\vk}-\frac{\Delta_{2}^{2}v_{\vk}^{2}}{\xi_{\vk}+\eta}
\approx{}E_{\vk}-v_{\vk}^{2}\Delta_{1}\zeta\label{eq:pathInt2:xiExpand2}\\
E_{3\vk}&\approx{}\xi_{\vk}+\eta-\frac{\Delta_{2}^{2}}{2(\xi_{\vk}+\eta)}
\approx{}\epsilon_{\vk}+\eta-\frac{\zeta}{2}\Delta_{1}
\label{eq:pathInt2:xiExpand3}
\end{align}
    $E_{1\vk}$ and $E_{2\vk}$ correspond to the traditional Bogoliubov quasi-particle modes; while $E_{3\vk}$ describes the fermionic excitation mostly in the closed channel.  The correction due to the inter-channel Pauli exclusion are all of order $\zeta$.

 \subsection{Renormalization of the mean-field equation}
    The summations in the gap equations above (Eq. \ref{eq:pathInt2:mf}) diverge when they are converted into integrals at 3D because of the artificial assumption of contact interactions. 
We can nevertheless remove the singularity in two steps.  First, we notice that the closed-channel bound state is much smaller than the interparticle distance. Therefore the two-body correlation within the closed channel is almost unchanged from its two-body value.  Given this consideration, we project the closed-channel correlation $h_{2\vk}$  onto the two-body  bound-state wave function of the uncoupled closed channel, $\phi$.  Second, the bare interaction in the open channel is replaced with a more physically meaningful quantity, the effective s-wave scattering length, $\tilde{a}_s$, which already incorporates the singularity. We briefly illustrate the procedure as follows, and please refer to \cite{Zhuthesis} for more details. 

 First, we project the closed-channel correlation $h_{2\vk}$ into  $\phi$
\begin{equation}\label{eq:pathInt2:hphi}
h_{2\vk}=\alpha\phi^{}_{\vk}u_{\vk}^{2}
\end{equation}
This projection is only for the high momentum, for the low momentum, the available phase space for the closed channel is much more limited because the atoms in the open channel center in the low momentum.  More specifically, this restriction is represented by the factor $u_{\vk}^2$.  We come to this conclusion because the closed-channel bound state $\phi$ is much smaller in size than the interparticle distance.  This guarantees that the low momentum states are dominated by the open channel.  
Comparing this equation with the two-body \sch equation for $\phi$, we find
\begin{equation}\label{eq:pathInt2:alpha}
\alpha=\frac{\sum_{\vk\vk'}{\phi_{\vk}^{*}}{Y_{\vk\vk'}}{h_{1\vk'}}}{\br{-E_{b}+\eta-2\mu-\lambda_{1}}}
\end{equation}
where
\begin{equation}\label{eq:pathInt2:lambda1}
\lambda_{1}(\eta)\equiv-\sum_{\vk}{\phi_{\vk}^{*}}{(E_{\vk}-\xi_{\vk})}\phi_{\vk}
	-\sum_{\vk\vk'}{\phi_{\vk}^{*}}{v_{\vk'}^{2}V_{\vk\vk'}}\phi_{\vk'}
\end{equation}

Put all these together into the gap equation of the open channel, we find 
\begin{equation*}
\begin{split}
\Delta_{1\vp}=&\sum_{\vk}\br{U_{\vp\vk}+
	\frac{\sum_{\vk^{'}\vp'} Y_{\vp\vp'}{\phi_{\vp'}}{\phi_{\vk'}^{*}}{Y_{\vk\vk'}}}
		{\br{-E_{b}+\eta-2\mu-\lambda_{1}}}}{{h_{1\vk}}}\\
	&-\frac{\sum_{\vk\vk{'}\vp'} Y_{\vp\vp'}{\phi_{\vp'}}{\phi_{\vk'}^{*}}{Y_{\vk\vk'}}v_{\vp'}^{2}{h_{1\vk}}}
		{\br{-E_{b}+\eta-2\mu-\lambda_{1}}}{}
\end{split}
\end{equation*}
Comparing to the similar equations in the two-body problem, we find the detuning from the resonance here differs from that of the two-body physics by $2\mu+\lambda_1$.  $2\mu$ describes the many-body shift of the starting point from zero to the Fermi surface, while $\lambda_1$ describes the inter-channel Pauli exclusion.  Furthermore, the last term in the r.h.s.  is also unique for the many-body problem.  This term has no singularity for high-momentum though because of the extra $v_{\vp'}^2$ term in the integral. 

Now we handle the above equation with  the similar strategy as in the single-channel problem: we integrate out the high-momentum part and replace the bare interaction with the more physically observable effective open-channel $a_s$.  
Multiply both side with $(1+TG)$,  where $T$ is the scattering matrix for the open channel, and $G=(\omega-H_{0})^{-1}$ is the Green's function for a free pair in the open channel.\footnote{Here we use the relation between the scattering $T$-matrix, the free pair Green's function $G=(\omega-H_{0})^{-1}$ and the bare interaction $V=-U_{\text{eff}}$.
\begin{equation*}
T=V+TGV=-(1+TG)U_{\text{eff}}
\end{equation*} }
\begin{equation*}
(1+TG)\Delta_{1}=-Th_{1}-\lambda_{2}
\end{equation*}
\begin{equation}\label{eq:pathInt2:lambda2}
\lambda_{2\vp}\equiv\alpha\sum_{\tilde\vp\,\vp'}(1+TG)_{\vp\tilde\vp} Y_{\tilde\vp\vp'}{\phi_{\vp'}}v_{\vp'}^{2}
\end{equation}

Using the zero energy value of the free pair Green's function $G(\omega=0)=(-2\epsilon_{\vk})^{-1}$ and introducing the low-momentum s-wave scattering length with extra shift 
\begin{equation}\label{eq:pathInt2:asKshift}
\tilde{a}_s=a_{\text{bg}}(1+\frac{\mathcal{K}}{\delta-2\mu-\lambda_{1}})
\end{equation}
where $\mathcal{K}=\Delta_{\mu}B$, $\Delta_{\mu}$ is the  effective difference of the pair magnetic moments of two channels. And  $\delta$ is the energy detuning between two channels, we have the renormalized open-channel gap equation
\begin{equation}
\begin{split}\label{eq:pathInt2:gapRenorm}
1=&-\frac{\lambda_{2}}{\Delta_{1}}\\
&-\mbr{\frac{4\pi{\tilde{a}_{s}(\mu,\lambda_{1})}}{m}\sum(\nth{2E_{\vk}}-\nth{2\epsilon_{\vk}}-\frac{\Delta_{2}^{2}\xi_{\vk}}{4(\xi_{\vk}+\eta){E_{\vk}^{3}}})}
	\\
\end{split}	
\end{equation}
Here the two factors $\lambda_1$ and $\lambda_2$, given respectively by Eqs. \eqref{eq:pathInt2:lambda1} and \eqref{eq:pathInt2:lambda2}, as well as the last term in the above equation, describe the inter-channel Pauli exclusion effect. All of them involve overlap integrals between the open-channel wave function and the closed-channel wave function.  (The factor $v_{\vk}^{2}$ or $u_{\vk}^{2}$ describes mostly the open-channel wave function, while $\phi$ describes the closed-channel wave function). The larger the overlap of the two, the larger are $\lambda_1$ and $\lambda_2$.  This has a very intuitive interpretation:  more overlap leads to more severe inter-channel Pauli exclusion, which in turn leads to larger (corrections) terms.  In our model, the open-channel wave function is  spread  over  a large region of  real space, (even on the BEC side, the real bound-state is very loosely bound comparing to the closed-channel bound state), while the closed-channel wave function is very sensitive to the binding energy, $E_{b}(\approx\eta)$.  A closed-channel bound state is more spread out  in real space and has larger overlap with the open-channel wave function,  if it is closer to the threshold, i.e.,  the binding energy  is smaller. Consequently,  the terms $\lambda_1$ and $\lambda_2$ are larger in such cases.                           Nevertheless, $\lambda_1$ is much smaller than the  Fermi energy $E_{F}$, or the other shift, the chemical potential, $2\mu$.  So it is still legitimate to treat this shift as a perturbation.  In addition, it can be shown that all these terms are linear in the density. (Please refer to \cite{Zhuthesis} for details.)

We can also use the expansion on $E_{i\vk}$ in Eqs. (\ref{eq:pathInt2:xiExpand}-\ref{eq:pathInt2:xiExpand3}) to rewrite the two number equations Eqs. (\ref{eq:pathInt2:numOpen}, \ref{eq:pathInt2:numClose}) to the first order in $\zeta$ 
\begin{gather}\label{eq:pathInt2:closeD2}
N_{\text{closed}}\approx\sum_{\vk}\frac{\Delta_{2}^2}{(\xi_{\vk}+\eta)(2\xi_{\vk}+\eta)}
\end{gather}
\begin{equation}\label{eq:pathInt2:openD2}
N_{open}\approx\sum_\vk\mbr{\frac{E_\vk-\xi_\vk}{2E_\vk}(1+\frac{\Delta_{1}}{\eta}\zeta)-\frac{\Delta_{1}^{3}}{4E_\vk^{3}}\zeta	}	
\end{equation}

Note that the closed-channel correlation $h_{2\vk}$ is forced into a particular simple form Eq. \ref{eq:pathInt2:hphi}, which does not lead to divergence in high momentum in integration and therefore has already been ``renormalized''.  This is  because the two-body closed-channel wave function $\phi$ is much smaller comparing to the inter-particle distance and therefore deformed little in many-body case to the lowest order  of $\zeta$. 
This equation implicitly includes $\Delta_2$ through $h_{2\vk}$ and therefore we do not need any equation explicitly about $\Delta_2$. 

In the summary, Eqs. (\ref{eq:pathInt2:hphi}, \ref{eq:pathInt2:gapRenorm}, \ref{eq:pathInt2:closeD2}, \ref{eq:pathInt2:openD2}) together become the renormalized set of equations that determine the state of the system at the mean-field level. 

%
%
%
%
%
%
%
%
\subsection{Discussion of the mean-field solution\label{sec:pathInt2:mean2}}
As discussed before, the correction of the narrow Feshbach resonance can be taken into account in two steps.  First,  omitting the inter-channel Pauli exclusion, we only consider the chemical potential $\mu$ in the shift and  the extra counting due to the closed channel.  Then in the second step, we can correct the previous result with quantities originated from the inter-channel Pauli exclusion unique to the three-species problem. 

In  the first step, the gap equation and the (open-channel) number equation are simplified to 
\begin{gather}
1=-\mbr{\frac{4\pi{\tilde{a}_{s}(\mu)}}{m}\sum(\nth{2E_{\vk}}-\nth{2\epsilon_{\vk}})}\label{eq:pathInt2:narrowGapS}\\
N_{\text{open}}=\sum_\vk\frac{E_\vk-\xi_\vk}{2E_\vk}\label{eq:pathInt2:narrowNumS}
\end{gather}
Here we only consider the shift of the  chemical potential $2\mu$ in   $\tilde{a}_s$ (Eq. \ref{eq:pathInt2:asKshift}),
\begin{equation}
\tilde{a}_{s}=a_{\text{bg}}(1+\frac{\mathcal{K}}{\delta-2\mu})\approx{}\frac{a_{\text{bg}}\mathcal{K}}{\delta-2\mu}
\label{eq:pathInt2:simplenarrowAs}
\end{equation}
The above equations need to be solved self-consistently. 

 An interesting point  to notice is that the gap $\Delta_1$ saturates in the BEC side. This is because the effective attraction in the open channel where the pairing happens is limited by the real attractive strength in the closed channel.  It can not become  infinitely strong as in an ideal single-channel model.  Mathematically, this can be seen from the gap equation, Eqs. (\ref{eq:pathInt2:mf}-\ref{eq:pathInt2:h2}). The closed-channel correlation is limited by the total density, and therefore the gap has a maximum.  

Once this step is finished, we can look for the correction due to the inter-channel Pauli exclusion.  Both $\lambda_1$ and $\lambda_2$ can be shown  to be linear with the density, so can be calibrated by experiments of different densities.  With the proper value of these two parameters, we can find the correction numerically.  It is nevertheless not hard to show that the correction is of order $\zeta$, so that we are warranted to  treat this effect as a perturbation \cite{Zhuthesis}.

\section{Beyond the mean-field: the collective modes\label{sec:bosonic}}
The order parameters ($\Delta_{1}$, $\Delta_{2}$) are defined in terms of the collective behaviors of many fermion atoms.  Fluctuations of the order parameters thus signal the collective excitation modes of the system. Here with a two-component complex order parameter, four independent modes exist:   two for the magnitude variation of each $\Delta_i$, the internal phase between $\Delta_1$ and $\Delta_2$, and the overall local phase $\theta(x)$ of $\Delta_1$ and $\Delta_2$.  
A similar result has been obtained in the BCS limit by Catelani and Yuzbashyan,\cite{Catelani}.
Two modes of the magnitude-fluctuation are gapped and massive as expected.  The modes of the phase-fluctuation are  of more interest.
We summarize our finding in the following.  Please refer to \cite{Zhuthesis} for more details.
\subsection{The in-sync phase mode\label{sec:insync}}
The in-sync phase mode is the counterpart of the Anderson-Bogoliubov modes in the single-channel problem\cite{RanderiaBEC, Nagaosa}. 
In this mode,  $\Delta_{1\,\vk}$ and $\Delta_{2\,\vk}$ rotate simultaneously, and the action $S(\bar{\Delta}_i,\Delta_i)$ (Eq. \ref{eq:pathInt2:actionD}), is invariant for such fluctuation.  
 We therefore conclude that there exists a massless (Goldstone) mode corresponding to the local phase invariance.  Introduce the phase fluctuation $\theta$, 
\begin{equation*}
\Delta_{i}(x)\rightarrow{}\Delta_{i}e^{i2\theta(x)}\qquad{}
\bar{\Delta}_{i}(x)\rightarrow{}\bar{\Delta}_{i}e^{-i2\theta(x)}
\end{equation*}
In order to focus on this particular mode, we  replace every  degrees of freedom except $\theta$ with their equilibrium values.  For small perturbation around the equilibrium, we only retain the first non-trivial order of $\theta$ (the second order).  It can then be shown that the fluctuation has a linear dispersion relation and therefore is a sound-like mode.   
\begin{equation}
S[\theta]=\sum_{q}\theta(q)\theta(-q)\big[\nth{2}\pi^{(0)}(0)\omega_m^2-\frac{n}{2m}\mathbf{q}^2\big]
\end{equation}
where $n$ is the density of the pairs and 
\begin{equation}
\pi^{(0)}(0)\equiv\sum_{k}\tr\br{G_{k}\sigma_3G_{k}\sigma_3}
\end{equation}
$G_{k}$ is the mean-field fermionic correlation in Eq. \ref{eq:nG} and $\sigma_3$ is defined as 
\begin{equation}
\sigma_3=\mtrx{1&0&0\\0&-1&0\\0&0&-1}
\end{equation}
Following the same approach as the last section, $G_{k}$ can be expanded with respect  to $\zeta$, after some algebra, we can write down the $\pi^{(0)}(0)$ to the first non-trivial order of $\zeta$.
\begin{equation}
\pi^{(0)}(0)\approx\sum_{\vk}\frac{\Delta_{1}^{2}}{E_{\vk}^{3}}
	-\sum_{\vk}\frac{\Delta_{1}^{2}\Delta_{2}^{2}\xi_{\vk}}{2E_{\vk}^{5}(\xi_{\vk}+\eta)}
\label{eq:pathInt2:pi0}
\end{equation}
$\Delta_1$ and $\Delta_2$ are the mean-field values of two order parameters. 
The first term in the above formula is the same as that of the single-channel crossover\cite{RanderiaBEC}.   It then becomes clear that the sound velocity follows the same structure as that for the single-channel crossover problem  with a correction of order  $\zeta$.
\subsection{The out-of-sync phase mode}

The out-of-sync phase mode on the other hand is a novel mode associated with the two-channel problem, with no direct counterpart in the single-channel problem.  
 When  the phase fluctuation of two channels are out of  sync,  the inter-channel coupling strength changes.  Thus, this mode is  expected to be  a gapped (massive) mode.  Similar as Sec. \ref{sec:insync}, we narrow down to the mode that the phases of two atoms  ($\psi_{b}$ and $\psi_{c}$) are opposite and leave all other degrees of freedom constant.  
\begin{equation*}
\mtrx{\psi_{a}(x)\\\psi_{b}(x)\\\psi_{c}(x)}\rightarrow{}
	\mtrx{\psi_{a}(x)\\\psi_{b}(x)e^{+i\theta(x)}\\\psi_{c}(x)e^{-i\theta(x)}}
\end{equation*}
The order parameters do not have a simple transformation because they are connected to two channels via $2\times2$ interaction matrix $\tilde{U}$, which mixes two channels (Eq. \ref{eq:pathInt2:DeltaPhi}).  
\begin{equation*}
\begin{pmatrix}\Delta_{1}(x)\\\Delta_{2}(x)\end{pmatrix}\rightarrow{}
	\mtrx{U&Y\\Y^{*}&V}\begin{pmatrix}\psi_{b}\psi_{a}(x)e^{+i\theta(x)}\\\psi_{c}\psi_{a}(x)e^{-i\theta(x)}\end{pmatrix}
\end{equation*}
This term cannot be easily written in terms of mean-field value $\Delta_i$.   On the other hand, as mentioned before, we freeze all  other modes to their mean-field value.  We therefore use another two-component of atom pairs $({\psi_{b}\psi_{a}},{\psi_{c}\psi_{a}})$, which is the linear recombination of $(\Delta_{1},\Delta_{2})$.  
   After some lengthy algebra, we can show that the action is 
\begin{equation}\label{eq:pathInt2:outofphase}
S[\theta]=\sum_{q}\theta(q)\theta(-q)\big[\nth{4}\pi^{(0)}(0)(\omega_m^2-\omega_{0}^{2})-\frac{n}{4m}\mathbf{q}^2\big]
\end{equation}
where
\begin{gather*}
\omega_{0}^{2}=-\frac{16Y\tilde{h}_{1}^{*}\tilde{h}_{2}}{\pi^{(0)}(0)}\\
 \tilde{h}_{1}=\av{\psi_{b}\psi_{a}}=\sum_{\vk}h_{1\vk}\qquad
\tilde{h}_{2}=\av{\psi_{c}\psi_{a}}=\sum_{\vk}h_{2\vk}
 \end{gather*}
Here we can see that  $\omega_{m}$ has a finite value $\omega_{0}$ at the zero momentum, which indicates a gapped mode. It can be shown that $\omega_{0}$
 is in the order of the ionization (pair breaking) energy, which is around $\Delta_1$ on the BCS side, while around $\abs{\mu}$ on the BEC side. 

Again, please refer to \cite{Zhuthesis} for more details. 
\section{Conclusion  and discussion\label{sec:conclusion}}
In this paper, we have studied the narrow Feshbach resonance in  the three-species case where two channels share the same species.  

In general, for the narrow resonance without a shared species, the main correction to the single-channel result comes from the  extra counting of the atoms in the open channel, which leads to the extra shift $2\mu$ in $\tilde{a}_{s}$, and the closed channel, which leads to the extra number equation.  Two number equations exist, one for each channel.  The open-channel number equation resembles the number equation of the single-channel model.

When there is a common species,  however, the  Pauli exclusion between two channels due to the common species in the three-species narrow resonance, calls for careful  consideration. 

  Our treatment follows the the  idea of ``universality'' \cite{Tan2008-1,shizhongUniv}.  For a dilute system with a short-range potential, such as the dilute ultracold alkali gas, the short-range part of a two-body correlation does not significantly change from two-body  to many-body.  This particular feature justifies using the two-body quantities  as the boundary condition for the many-body correlations.  In non-resonant situations, the s-wave scattering length, $a_s$, is a good candidate for such a purpose.  However, in a Feshbach resonance, the open-channel scattering length becomes enormous, comparable to or even larger than the many-body length scale; furthermore, its value is very sensitive to the ratio of weights of atoms in the two channels.  Therefore it is no longer suitable as the two-body characteristic quantity used in  the many-body problem. A better candidate of two-body quantities as boundary condition here is the short-range part of the two-body wave function itself.  In other word, we expect the two-body correlation of the many-body system matches the shape of the two-body wave function  at short distances.

When the spatial extension $a_c$ of  the closed-channel bound state is  of the order of the   inter-particle distance, $a_{0}$, or even larger, the Feshbach resonance in the many-body context is a genuine three-species many-body problem and no simple solution is available to our knowledge.  By contrast, when the bound-state's spatial extension is much smaller than the interparticle distance, the two-body correlation in the closed channel, which is expected to almost entirely concentrated in the short-range part,  is proportional to  the two-body bound-state wave function. The ratio of the the bound-state size and the interparticle distance, $a_{c}/a_{0}\sim\zeta$, serves as the expansion parameter and we can extract  the effect of the inter-channel Pauli exclusion perturbatively.  In essence, we can then  ignore the many-body effects within  closed-channel bound states, while only taking into consideration of the Pauli exclusion between channels and within the open channels.  A few new parameters need to be introduced and can be calibrated from experiments, such as $\lambda_{1}$, $\lambda_{2}$.  Mean field properties can still be determined through gap equations and number equations similar to the single-channel case.  The excitation modes are also close to the original single-channel result with correction of the order of $\zeta$.

     In our approach, we take the broad resonance result (or the single-channel crossover) as our zeroth order solution, upon which the expansion is performed.  It is however known that the simple BCS-type pairing treatment is not adequate  to quantitatively describe the whole BEC-BCS crossover region.  Therefore the zeroth order solution used here can be improved through further theoretical development.  Nevertheless, we expect the perturbation approach used here to build the narrow resonance result from the single-channel crossover result to remain valid.  Once the zeroth order solution (for a broad resonance or a single channel BEC-BCS crossover model) is appropriately improved, the correction of the narrow resonance in such a parameter regime, can still be obtained by  a procedure similar to the one described in this paper.

\begin{acknowledgements}
We thank  Professor Monique Combescot, Dr. Shizhong Zhang and  Dr. Wei-Cheng Lee for many inspiring discussions. Part of this research  is supported  by the National Science Foundation under grant No. DMR 09-06921. 

\end{acknowledgements}

\appendix

\section{Diagonalization of  the mean-field Green's function Eq. (\ref{eq:nG})\label{sec:diagonalize}}
 Apply $T$ onto $G^{-1}$ (Eq. \ref{eq:nG}), we have 
\begin{equation}\label{eq:pathInt2:G2}
T_{\vk}^{\dg}G_{\omega_{n},\vk}^{-1}T_{\vk}=i\omega_nI+\mtrx{-E_{\vk}&0&u_{\vk}\Delta_2\\0&+E_{\vk}&v_{\vk}\Delta_2\\u_{\vk}\Delta_2&v_{\vk}\Delta_2&+\xi_{\vk}+\eta}
\end{equation}
We drop all the $k$ subscripts in the rest of this section because matrices in this section are decoupled in momentum and we only deal with one particular momentum $\vk$ a time. The off-diagonal elements in the above matrix is regarded as perturbation because  we  only seek the solution around the BCS wave function ($T$ transformation). 
We need to find a unitary transformation $L$ to diagonalize this  matrix.
 We notice that the first term is proportional to an identity matrix and does not change by unitary transformation, we only need to concentrate on the second term.  We rescale all elements with $E_{\vk}$ for simplicity in the following of this section. 
\begin{equation*}
y=\frac{\Delta_2}{E_{\vk}},\qquad
 t=\frac{\xi_{\vk}+\eta}{E_{\vk}},\qquad
\end{equation*}
 And the second term in r.h.s. of Eq. \ref{eq:pathInt2:G2} is 
\begin{equation*}
R=
\begin{pmatrix}
-1&0&uy\\
0&1&vy\\
uy&vy&t
\end{pmatrix}
\end{equation*}
The secular equation of $R$ is ($\abs{x\,I-R}=0$)
\begin{equation}\label{eq:pahtApp:secular}
(x^{2}-1)(x-t)-y^{2}x+(u^{2}-v^{2})y^{2}=0
\end{equation}
We use $u^{2}+v^{2}=1$ here.  We  assume at the zeroth order, the three eigenvalues are $-1$, $1$ and $t$.  ($t$ has weak dependency on energy as $(\xi_{k}+\eta)/E_{k}$, however, at the low energy region of interest, we ignore $\xi_{k}$.) Both $y$ and t are larger than 1, however, we will verify that given condition $y^{2}\ll{t}$, the correction is indeed small and the expansion is reasonable (See Appendix \ref{sec:pathApp:consistency}).  \emph{Indeed,  close-channel component can still be smaller than the open-channel component at low-k (in the order of $k_{F}$)  due to the close-channel bound state is much smaller than the interparticle distance even when the total close-channel atom number  is more than that of the open channel. }  And here all the quantities are about low-k unless specifically noticed.
We expand the system to the first order of the dimensionless parameter $\tilde\zeta=y^{2}/{t}$ (\ref{eq:pathInt2:zetaDef})
, and find
\begin{equation}
\begin{array}{ccc}
x^{(0)}&\quad{}x^{(1)}&\quad{}Eigenvector\nonumber\\
-1&-u^{2}\tilde\zeta&\mtrx{1&\frac{uvy^{2}}{2t}&-\frac{uy}{t}}\\
1&-v^{2}\tilde\zeta&\mtrx{-\frac{uvy^{2}}{2t}&1&-\frac{vy}{t}}\\
t&\nth2\tilde\zeta&\mtrx{\frac{uy}{t}&\frac{vy}{t}&1}
\end{array}
\end{equation}
Now it is easy to write down the corresponding diagonal matrix and the unitary transformation
\begin{equation}
B=i\omega_{n}I+E\mtrx{-1-u^{2}\tilde\zeta&0&0\\0&1-v^{2}\tilde\zeta&0\\0&0&t+\nth2\tilde\zeta}
\end{equation}
\begin{equation}
L=\mtrx{1&-\frac{uvy^{2}}{2t}&\frac{uy}{t}\\\frac{uvy^{2}}{2t}&1&\frac{vy}{t}\\-\frac{uy}{t}&-\frac{vy}{t}&1}
\end{equation}
Here $L$ is not exactly unitary transformation, it is only unitary in the first order of  $\tilde\zeta$. We have 
\[
B=i\omega_{n}I+E\,(L^{\dg}RL)+o(\tilde\zeta)
\]
Restore the factor $E_{\vk}$ and we can obtain Eq. (\ref{eq:pathInt2:L1}) and Eq. (\ref{eq:pathInt2:Bapprox}).
(Here we use $uv=\Delta_{1}/2E$.)

In the above treatment, the small parameter $\tilde\zeta$ is momentum dependent.  If we restore the subscript $\vk$ and scale it  back with $E_{\vk}$
\begin{equation}
\tilde\zeta=\frac{\Delta_{2}^{2}}{E_{\vk}(\xi_{\vk}+\eta)}
\end{equation}
A momentum-dependent small parameter is not very convenient to work with, so we take its maximum value in low momentum ($\lesssim{}E_{F}$).  In the BCS-like states ($\mu>0$), $\min{E_{k}}=\Delta_{1}$, $\min{\xi_{\vk}}=0$; in the BEC-like states ($\mu<0$), $\min{E_{k}}=\sqrt{\Delta_{1}^{2}+\mu^{2}}$ and $\min{\xi_{\vk}}=\abs{\mu}$. We take the smaller values and have our expanding small parameter $\zeta$(\ref{eq:pathInt2:zetaDef})
\begin{equation}
\zeta=\max\tilde{\zeta}=\frac{\Delta_{2}^{2}}{\Delta_{1}\eta}
\end{equation}

\section{Smallness of the expansion factor $\zeta$\label{sec:pathApp:consistency}}
Here we check the smallness of our expansion factor $\zeta$(\ref{eq:pathInt2:zetaDef}).  We have the closed-channel gap equation (\ref{eq:pathInt2:mf})
\begin{equation}
\Delta_{2}=\sum{}Yh_{1\vk}+\sum{}Vh_{2\vk}\label{eq:pathInt2:mfclose}
\end{equation}
The first term on the right is relatively small comparing to the second term.  Therefore, we drop the first term in estimation.  Furthermore,  we assume $h_{2\,\vk}=\sqrt{N_{c}}\phi_{\vk}$, where $N_c$ is the total number of closed-channel pairs, and $\phi_{\vk}$ is the normalized wave function of the  isolated closed-channel potential satisfying the two-body \sch equation 
\begin{equation}\label{eq:pathInt2:phi}
-E_{b}^{(0)}\phi_{\vp}=2\epsilon_{\vp}\phi_{\vp}-\sum_{\vk}V \phi_{\vk}
\end{equation}
Rearranging it, we have (especially at low momentum)
\begin{equation*}
\sum_{\vk}V \phi_{\vk}=(2\epsilon_{\vp}+E_{b})\phi_{\vp}\approx{\eta}\phi_{\vp}
\end{equation*}
Here $E_{b}$ is the binding energy of the closed-channel bound state, which is close to the Zeeman energy difference, $\eta$, around the Feshbach resonance. The second approximation is correct at low momentum not too far away from the resonance (smaller or in the same order of the Fermi momentum), i.e. $\epsilon_{\vp}\ll{}E_{b}\approx\eta$ .  Put all these together, we have
\begin{equation*}
\Delta_{2}\approx\alpha{}E_{b}\phi_{k=0}
\end{equation*}
If we assume a simple exponentially decayed wave function:
\begin{equation}\label{eq:pathInt2:phi2body}
\phi_{\vk}=\sqrt{\frac{8\pi\kappa}{\mathcal{V}_{0}}}\frac{1}{k^{2}+\kappa^{2}}\approx\sqrt{\frac{8\pi\kappa}{\mathcal{V}_{0}}}\frac{1}{\kappa^{2}}
\end{equation}
Here  $\mathcal{V}_{0}$ is the total volume and $\kappa$ is the characteristic momentum of the closed-channel bound state, $\eta\approx{}E_{b}=\hbar^{2}\kappa^{2}/2m$.  The second approximation above is only for  low momentum.  Collect all these together, we have
\begin{equation}
\Delta_{2}\approx\sqrt{N_{c}}\eta\sqrt{\frac{8\pi\kappa}{\mathcal{V}_{0}}}\frac{1}{\kappa^{2}}
\sim\eta\br{\frac{k_{Fc}}{\kappa}}^{\frac{3}{2}}
\end{equation}
$k_{Fc}\sim(N_c/\mathcal{V}_0)^{\nth{3}}$ is the Fermi momentum corresponding to the density of the closed-channel pairs, which is much smaller than the characteristic momentum for the bound-state, $\kappa$.  

Now let us get back to $\zeta$ (\ref{eq:pathInt2:zetaDef})
\begin{equation}\tag{\ref{eq:pathInt2:zetaDef}}
\zeta=\frac{\Delta_{2}^{2}}{\Delta_{1}\eta}
\end{equation}
At the  BCS limit, the closed-channel density is small, $k_{F\,c}$ is small and that makes $\zeta$ small; Moving toward the (narrow) resonance, where the closed-channel density is comparable to to the total density, at low energy, $\Delta_{1}$ is in the order of the Fermi energy.   We have (we no longer distinguish $k_{F\,c}$ with $k_{F}$)
 \begin{equation}\label{eq:pathApp:zetaEs}
 \zeta=\frac{\Delta_{2}^{2}}{\Delta_{1}\eta}\sim\frac{\eta^{2}\frac{k_{Fc}^{3}}{\kappa^{3}}}{k_{F}^{2}\eta}\sim\frac{k_{F}}{\kappa}\ll1
\end{equation}
which is also very small. 
\bibliography{citation}

\end{document}